\documentclass[onecolumn,showpacs,preprintnumbers,amsmath,amssymb,12pt]{revtex4}

\usepackage{graphicx}
\usepackage{dcolumn}
\usepackage{bm}

\begin{document}

\preprint{APS/123-QED}

\title{The emergence of prime numbers as the result of evolutionary strategy}

\author{Paulo R. A. Campos}\email{prac@ifi.unicamp.br}\thanks{Permanent Address: Departamento
de F\'{\i}sica e Matem\'atica, Universidade Federal Rural de Pernambuco, Dois Irm\~aos 52171-900,
Recife-PE, Brazil}
\author{Viviane M. de Oliveira}
\author{Ronaldo Giro}
\author{Douglas S. Galv\~ao}

\affiliation{%
Departamento de F\'{\i}sica Aplicada, Instituto de F\'{\i}sica Gleb Wataghin, Universidade Estadual de Campinas 
13083-970 Campinas SP, Brazil
}%

\begin{abstract}
 We investigate by means of a simple theoretical model the emergence of prime 
numbers as life cycles, as those seen for some species
of cicadas. The cicadas, more precisely, the Magicicadas spend most of 
their lives below the ground and then emerge and die in a short 
period of time. The Magicicadas display an uncommon behavior: their emergence is synchronized and these periods 
are usually prime numbers. In the current work, we develop a spatially extended model at which preys 
and predators coexist and can change their evolutionary dynamics through the occurrence of
mutations. We verified that prime numbers as life cycles emerge as a result of the evolution 
of the population. Our results seem to be a first step in order to prove that 
the development of such strategy is selectively advantageous, especially for those organisms 
that are highly vulnerable to attacks of predators. 
\end{abstract}

\maketitle 

\newpage

It is well established that species evolve by increasing their adaptation to the environment where they live.
In this sense, Sewall Wright created the metaphor of an uphill climbing to the Darwinian evolution \cite{wright}.
In order to become selectively stronger, species improve 
their fitness through the occurrence of mutations at the genotype level which confers a selective advantage 
or by developing new strategies of competition with other species.

In this contribution we investigate, within a theoretical framework, the occurrence of 
periodical behavior for the life cycles in nature. Especially, we focus our analysis on finding mechanisms that
can generate life cycles which are prime numbers such as those known for the cicadas. 
The cicadas have attracted the attention 
of the scientific community since a long time ago
which dates from the end of the nineteenth century \cite{butler1886}. This great interest owes to the 
uncommon behavior displayed by those insects which is not found in any other species in nature. 
Despite the long period of investigation, the dynamics of the periodical cicadas is still poorly understood. 
The cicadas, more specifically the genus Magicicadas, have 13- or 17-year life cycles which are 
the longest life cycle known for any insect. The Magicicadas spend most of their 
lives underground before they emerge and assume their adult form, reproducing and dying 
within few weeks. Interestingly, all the cicadas in a given location emerge at the same time, in an impressive
synchronism \cite{lloyd1966,marlatt1907,kritsky89,karban80}. In different regions, the broods of cicadas can be out 
of synchrony \cite{kritsky89,simon89}. The most interesting feature of the life cycle of the Magicicadas
is that they appear in prime numbers, and so the cicadas have been pointed out as a biological generator 
of prime numbers \cite{markus2000,markus2001}.
Do the long life cycle and the fact they appear in prime number have evolutionary implications?
It has been proposed \cite{lloyd1966} that this could be the result of an evolutionary strategy 
to avoid parasites. It would be very difficult to the parasites to match the life cycle if it 
appears in prime numbers. For instance if the cicadas have a life cycle of 17 years and the
parasites of a couple of years they would meet only one time out of hundreds of years. This would
lead to the extinction of the parasites if they depend on the cicadas to reproduce. In fact no
specific periodic predator to cigadas have been found up to date.

Another relevant feature about the Magicicadas is their abundance when 
compared to other kind of cicadas \cite{may1979,hoppensteadt1976}. This abundance
enables to satiate predators and so to avoid extinction since the Magicicadas are extremely 
vulnerable to their predators, especially birds \cite{butler1886,beamer1928,lloyd1966}. 
As pointed out by Lloyd and Dybas \cite{lloyd1966}, the Magicicadas are really an 
unique phenomenon in Biology.

Few models have been reported addressing this unusual magicicada behavior, but a coherent 
theory to describe the evolutionary mechanism that guided the 
emergence of this synchronism and long life cycles is still missing. As far 
as we know the most relevant 
mathematical description of the cicadas behavior was developed by 
Hoppensteadt et al. \cite{hoppensteadt1976}. They present a mathematical
formulation for the cicadas which invokes the predator-prey relation and also consider the limiting capacity of the environment. 
As result, they showed that it is possible to find a synchronized and long life cycle solution for the cicadas if the 
system satisfy a set of conditions based on the parameters of the model.

In the current paper, we wish to investigate the appearance and advantages of producing evolutionary strategies that yield 
life cycles which are given by prime numbers. For that purpose, we introduce a spatial model where the 
agents can change their strategies by mutations which alter the length of their life cycles.
In our formulation, we use the cellular automaton approach to describe the spatial-temporal 
evolution of the population \cite{wolfram1986}. A cellular automaton is a regular spatial lattice of cells, each of these cells can assume
any one of a finite number of states. The state of each cell is updated simultaneously and the state of the 
entire lattice advances in discrete time steps. The state of each cell $\mathbf{s}(t+1)$ at time $t+1$ is determined by the state of its neighboring cells 
at the previous time $t$ according to a local rule.

Our approach resembles the predator-prey model \cite{lotka,voterra} and 
we only make use of local dynamics rules to evolve the population. 
Spatially extended models have been largely employed to study ecological models \cite{tilman97}, and provide a good way to 
explain deviations from the Lotka-Voterra dynamics. Besides, the emergent patterns of evolution in 
several systems can contribute with new insights that are not captured by the quantitative analysis 
which deals with differential equations and assumes homogeneous environments. As an example, we cite the spatial structure 
of evolution in prebiotic scenarios \cite{boerlijst}, where the 
emergent structural pattern shows to be a relevant
mechanism to ensure the maintenance of evolutionary information, and the outcome against 
parasites which invades the system. This formulation has the benefit that we do not have to work 
with a large set of parameters.

Although a simple spatially extended model has been proposed recently to describe the cicadas behavior \cite{markus2001}, the formulation 
considers that the predators also exhibit
a periodical life cycle and have similar dynamics as those of the preys. In this manner, the selection of life cycles concerns the optimal way to
make the emergence of preys do not coincide with the emergence of predators. However, there is no evidence for the existence of such kind of parasitoids in nature. 
It looks more realistic to assume the existence of a predator, for instance birds, which are constantly available to feed on the cicadas. Thus, our formulation is completely general and
address more fundamental questions rather than the production of life cycles which does not match with the life cycles of some sort of parasitoids.

Our model is defined as follows. We consider a two-dimensional lattice of linear size $L$ and $N = L \times L$ sites with periodic boundary conditions.
Each lattice site $\mathbf{s}_{i}$ can take one of the three possible states: $\mathbf{s}_{i}=0,1$ and $2$. The state $0$ denotes that the site is empty, the 
state $\mathbf{s}_{i}=1$ corresponds to a cell which is occupied by a prey, whereas state $\mathbf{s}_{i}=2$ means that a predator exists in that site.
To each prey, we ascribe a quantity $T_{inc}(i)$ which defines the period that it remains below the ground before emerging, i.e., the sequence of
events from the egg to the reproducing adult. Initially its values are randomly assigned in a pre-defined range.
As the system evolves in time these values can change induced by mutations. In the same way, we ascribe to each
predator a quantity $T_{starv}$ which defines the maximum time that it can remain alive without food supply. After that 
period of starvation the predator dies. We consider the Moore neighborhood, i.e., each cell interacts with its eight nearest neighbor cells (see {\it Figure 1}). 
The population evolves according to the following dynamical rules: 
\begin{itemize}
\item a cell in state $0$:
\begin{itemize}
\item [(a)] can change to state $1$ if there are at least $k_{prey}$ emerging neighbors cells in state $1$. In this case, we randomly
select a prey among those in the neighborhood to produce an offspring and to occupy the cell. The offspring inherits the same period of incubation $T_{inc}(i)$ of its
parent when mutations do not take place. If the offspring is hit by a mutation, which occurs with probability $U_{inc}$, the quantity $T_{inc}(i)$ equally
decrease or increase by one unit.
\item [(b)] it will remain in state $0$ if there are less than $k_{prey}$ neighbors in state $1$.
\end{itemize} 
\item a cell in state $1$:
\begin{itemize}
\item [(c)] During emergence, the cell can change to state $2$ if there are at least $k_{predator}$ predators in its Moore neighborhood. In this situation, we 
randomly choose a predator among those in its neighborhood to produce an offspring and to occupy the cell. The offspring 
inherits the same period of incubation $T_{starv}(i)$ of its parent when mutations do not take place. 
If the offspring is hit by a mutation, which occurs with probability $U_{starv}$, the quantity $T_{starv}(i)$ equally
decrease or increase by one unit.
\item [(d)] During emergence, if there are at least one predator and less than  $k_{prey}$ predators, the prey is eaten 
and the cell will be empty in the next generation, i.e., it will move to state $0$. 
\item [(e)] In all other situations, the cell will keep in state $1$.
\end{itemize}
\item a cell in state $2$: 
\begin{itemize}
\item [(f)]can change to state $0$ if for a time interval $T_{starv}(i)$, the predator haven't eaten any prey. 
\item [(g)] Otherwise, it will remain in state $2$. 
\end{itemize}
\end{itemize}
The mutation mechanism permits that the life cycles of the preys change by one unit at each occurrence and 
the predators change their period of resistance to the lack of food. In this way, the population generates a greater diversity of species, which
 enables the species to search for better evolutionary strategies. In this direction, previous investigations have 
demonstrated the occurrence of increasing and/or decreasing of the life cycles in Magicicada records \cite{kritsky89,simon89}.

In {\it Figure 2} we show the results for the distribution of the dominant value of life cycles of preys 
in the population after the steady regime is attained . 
By dominant we mean the most frequent value of $T_{inc}(i)$ in the population of preys. 
Although there is a dispersion of life cycles of preys in the population, which is 
higher for higher values of probability $U_{inc}$, the distribution of life cyles
 of preys is really peaked at some dominant prime-number value. We estimated the distribution 
from $1000$ distinct runs. In these simulations we considered a two-dimensional lattice of linear size $L=100$. 
From the Figure, we clearly can see that the prime life cycles for 
the preys dominates the distribution. We also notice that life 
cycles of small length are most likely to occur than long life cycles. On the other 
hand, the histogram for the life cycles of predators  does not 
display any particular pattern and we corroborated that the distribution is rather uniform (data not shown). This is a strong evidence of evolutionary advantages for the prime life cycles.
If this is true we can expect that if you increase the evolutionary pressure increasing the
numbers of predators the prime number strategy would be more evidenced and longer cycles shoud appear.

In {\it Figure 3a} we show the results for the distribution of life cycles of preys, now with the initial fraction of
predators on the lattice higher than in the previous simulations. 
As we expected the life cycles of preys are predominantly
prime numbers as before, and longer life cycles have higher chance to occur than in the previous cases shown in {\it Figure 2}.
In {\it Figure. 3b} we show the equivalent results for the predators. As we can
see from the figure prey and predators present explicitly a clear
differentiated strategies as they co-evolve. In contrast with the
preys no specific pattern associated with prime numbers was observed
for the predators.

From our simulations, we found out that a higher concentration of predators 
will usually come out a higher concentration of predators in
the long term evolution if the system maintains the coexistence of both predators and preys. 
Thus, the preys are effectively subjected to a stronger 
competition with their predators. In this manner, our results show that the preys tend to increase
their life cycles in such way that it allows them to avoid extinction, since the preys are extremely vulnerable to
the presence of predators. We have also ascertained that extinction of preys with 
the subsequent extinction of predators happens when the population of
predators attains high concentration values and so the predators feed on the preys as much as they can. 

From the evolution of spatial patterns in our simulations we detected some
kind of subpopulation segregations. This is in agreement with experimental
observations \cite{kritsky89,simon89}. For each subpopulation we do not
necessarily have an unique value of life cycle, but we observed some 
fluctuations centered around a dominant value. The same phenomenon also occurs
for the distribution of life cycles for the whole population.

In {\it Figure 4} we depict a typical scenario for the temporal evolution of 
the concentration of preys in a given population. We observe that the 
stationary state for the dominant life cycle of preys is attained 
in a few generations. In all our simulations, the time needed to reach the stationary
 value for the life cycle of preys is not greater than 5000 generations.
From the figure, we can see 
that while the predominant life cycle of preys is not a prime number, 
the density of preys decreases as the population evolves in time. The 
selection of a prime number for the predominant life cycle 
prevents a further decreasing of the density of preys in the population. 
This is a clear evidence that the selection of prime numbers of life 
cycles corresponds to the optimal strategy for the preys in order to
prevent extinction and to make the density grows. We have also observed 
that preys try to extend their life cycles as the density of predators increases  
but an indefinite growth of predators density leads to the collapse of the 
population with the resulting extinction of both species.
One interesting question is why $T_{starv}$ going to infinity is not a
good strategy for predators. In principle we could expect that this
could make them very robust to all prey strategies. However, if the predator
population increases beyond a certain limit the prey population 
would be drastically affected and could be extinct. This would also lead
the predators to extinction by starvation. Increasing $T_{starv}$
beyond a certain limit is equivalent to increase predation pressure
mentioned above with the resulting effect of extiction of both species.

In summary we have demonstrated based on a spatially extended cellular 
automaton model that the appearance of life cycle in prime
 numbers can be explained as a result of a winning 
evolutionary strategy of prey-predator competition. The 
prime life cycle dominates the population distribution and 
stabilizes the prey populations. Increasing predator competition 
favors longer prime cycles but after a certain limit of predator 
populations both species become extinct, which takes place when we consider
small $K_{predator}$
 values and/or high initial concentration of predators. Outside these limits 
our main conclusions are not set parameter dependent and the methodology is 
very robust. This is the first neutral and completely general model to 
demonstrate that the appearance of prime numbers in nature can be the
 result of a winning evolutionary strategy of prey-predator games.

\bigskip
\bigskip

The authors acknowledge the financial support of the Brazilian Agencies FAPESP and CNPq.

\newpage

\begin{figure*}[t]
\includegraphics[height=5.5cm,width=10.5cm,angle=0]{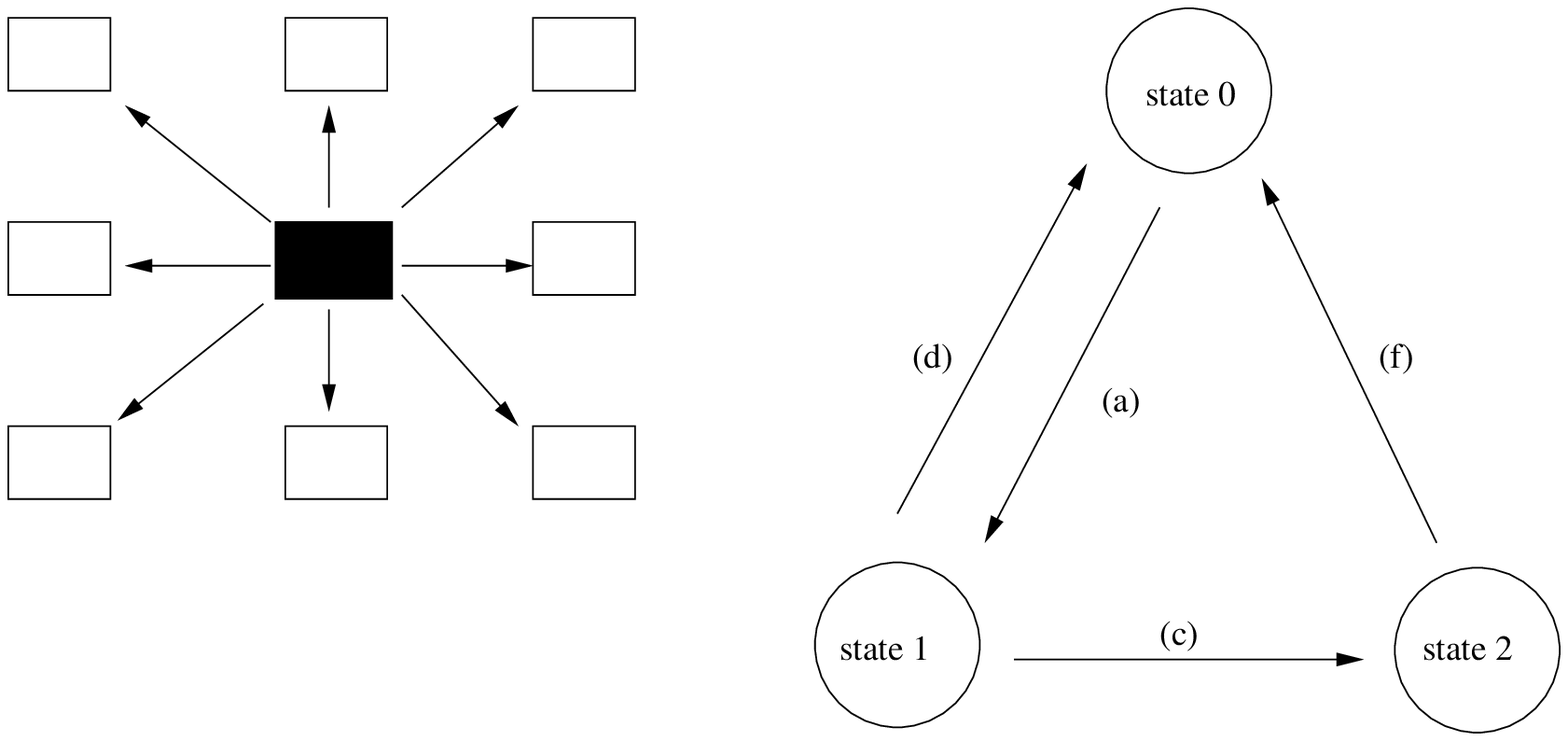}
\caption{\label{fig:figure1} Scheme of the dynamical rules of the cellular automaton. In part (a) we show the Moore neighborhood of a given cell. 
In part (b) we show the allowed transitions among the three states which is depicted 
by the directed arrows.}
\end{figure*}

\vspace{4cm}
\pagebreak

\newpage
\begin{figure}[t]
\includegraphics[width=8cm,angle=0]{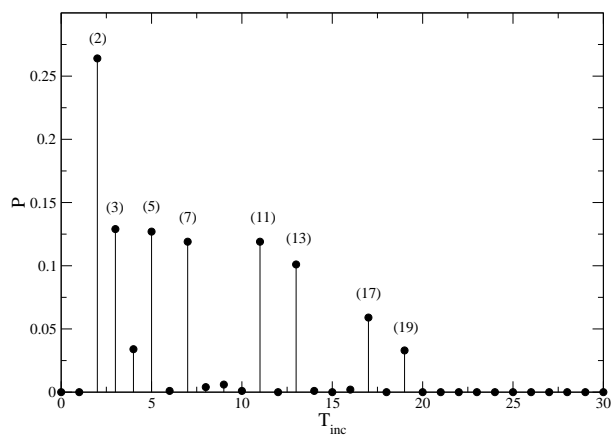}
\caption{\label{fig:figure2} Distribution of the dominant value of life cycles of preys after the population has evolved for $20,000$ generations.
The data were taken over $1000$ independent simulations. The parameter values are $L=100$, $k_{prey}=4$, $k_{predator}=4$, $U_{inc}=10^{-5}$
and $U_{starv}=10^{-5}$. The initial concentrations of empty sites $x_{0}$, preys $x_{prey}$ and predators $x_{pred}$ was $0.5$, $0.45$ and $0.05$, respectively. 
In all simulations we randomly assigned the initial values for $T_{inc}(i)$ and $T_{starv}(i)$.}
\end{figure}

\newpage

\begin{figure}[t]
\includegraphics[width=8cm,angle=0]{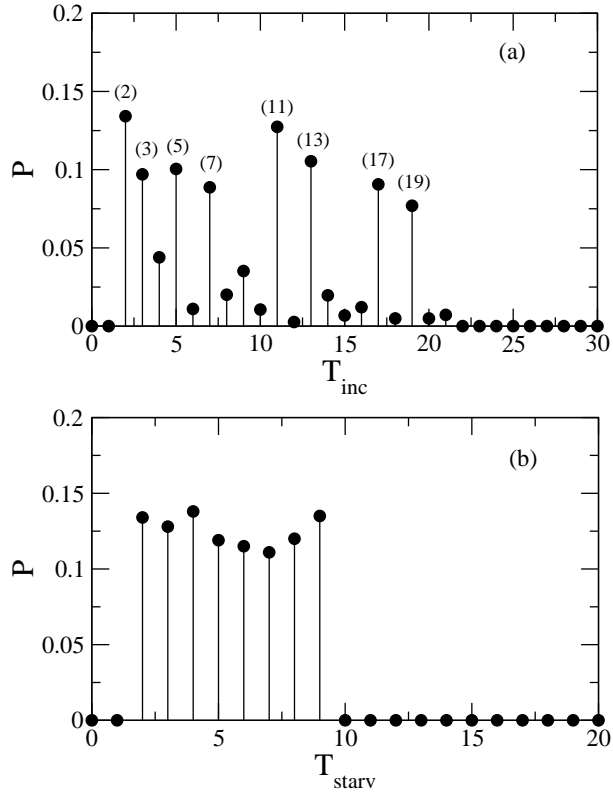}
\caption{\label{fig:figure3} Distribution of the dominant value of life cycles of (a) preys and (b) predators after the population has evolved for $20,000$ generations.
The data were taken over $1000$ independent simulations. The parameter values are $L=100$, $k_{prey}=4$, $k_{predator}=4$, $U_{inc}=10^{-5}$
and $U_{starv}=10^{-5}$. The initial concentrations of empty sites $x_{0}$, preys $x_{prey}$ 
and predators $x_{predator}$ is $0.5$, $0.4$ and $0.1$, respectively. 
In all simulations we randomly assigned the initial values for $T_{inc}(i)$ and $T_{starv}(i)$.}
\end{figure}

\newpage

\begin{figure}[tb]
\includegraphics[width=8cm,height=6cm,angle=0]{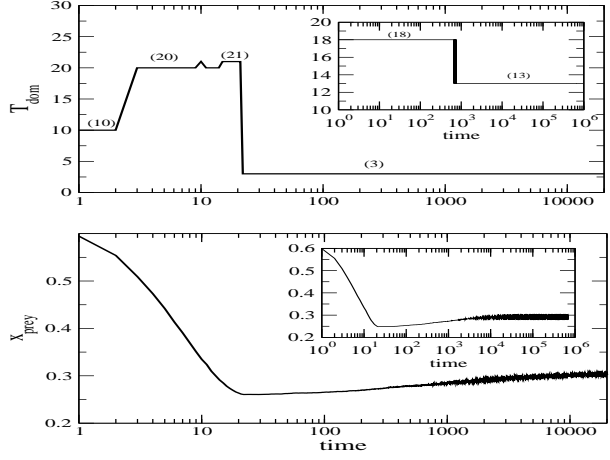}
\caption{\label{fig:figure4} Temporal evolution of the dominant value of life cycles of preys in the population, $T_{dom}$, and the 
concentration of preys. The parameters are $L=100$, $k_{inc}=k_{pred}=4$, $U_{inc}=10^{-5}$, $U_{starv}=10^{-5}$ 
and initial concentrations $x_{empty}=0.3$, $x_{prey}=0.595$ and $x_{predator}=0.105$. The numbers
between parenthesis in part (a) are the corresponding values of $T_{dom}$. In the insets we show the results for another run with 700,000 generations.
As we can see the results are qualitatively identical and demonstrate
that they are not number generation dependent after the steady
state is attained. }
\end{figure}

\end{document}